%% file: thesis.tex
\documentclass[12pt,a4paper]{report}

\usepackage{thesis}
\usepackage{float}
\usepackage{enumerate}
\usepackage{lmodern}
\usepackage[numbers,sort&compress]{natbib}
\usepackage[hidelinks]{hyperref}
\usepackage[left=2.5cm,top=2.5cm,right=2.5cm,bottom=2.5cm]{geometry}
\usepackage{nomencl}
\makenomenclature

\usepackage{graphicx}
\usepackage{xcolor}
\usepackage{amssymb}
\usepackage{amsthm}
\usepackage{mathtools}
\renewcommand{\paragraph}[1]{\smallskip\noindent\textbf{\emph{#1.}}}

\newtheorem{theorem}{Theorem}

\usepackage{tocloft}

\newcommand{\thesistitle}{A Game-theoretic Approach\\
\vspace{-4mm}for Provably-Uniform Random Number Generation\\in Decentralized Networks}
\newcommand{\thesisauthor}{Zhuo Cai}
\newcommand{\programname}{Computer Science and Engineering}
\newcommand{\departmentname}{Department of Computer Science and Engineering}
\newcommand{\thesisdate}{July 2023}
\newcommand{\signdate}{July 2023}
\newcommand{\thesisdegree}{MPhil}
\newcommand{\dedicate}{\textit{To Chang Lu}, my lovely and supportive mom. }
\newcommand{\supervisorinfo}{Prof. Amir Kafshdar Goharshady, Thesis Supervisor \\ Department of Computer Science and Engineering\\ Department of Mathematics}
\newcommand{\depheadinfo}{Prof. Xiaofang Zhou \\ Head, Department of Computer Science and Engineering}

\begin{document}

\pagenumbering{roman}
\pagestyle{plain}
\setcounter{page}{1}
\addcontentsline{toc}{chapter}{Title Page}
\include{1_title}

\newpage
\addcontentsline{toc}{chapter}{Authorization Page}
\include{2_authorization}

\newpage
\addcontentsline{toc}{chapter}{Signature Page}
\include{3_signature}

\newpage
\include{4_dedicate}

\newpage
\addcontentsline{toc}{chapter}{Acknowledgments}
\include{5_acknowledgement}

\newpage
\addcontentsline{toc}{chapter}{Table of Contents}
\tableofcontents

\newpage
\addcontentsline{toc}{chapter}{List of Figures}
\listoffigures

\newpage
\addcontentsline{toc}{chapter}{Abstract}
\include{abstract}

\newpage
\pagenumbering{arabic}
\pagestyle{plain}
\setcounter{page}{1}
\include{chapter1}
\include{chapter2}
\include{chapter3}
\include{chapter4}
\include{chapter5}
\include{chapter6}

\newpage
\addcontentsline{toc}{chapter}{References}
\bibliographystyle{IEEEtranN}
\include{reference}

\newpage
\addcontentsline{toc}{chapter}{Publications}
\include{publication}

\end{document}

%% file: 1_title.tex
\thispagestyle{empty}
\null\vskip0.5in
\begin{center}
  \begin{LARGE}
    A Game-theoretic Approach\\
    \vspace{-5mm}
    for Provably-Uniform Random Number Generation\\
    in Decentralized Networks
  \end{LARGE}
  \vfill
  \vspace{20mm}

  by

  \vspace{4mm}

  \thesisauthor \\
  \vfill
  \vspace{20mm}

  A Thesis Submitted to\\
  The Hong Kong University of Science and Technology \\
  in Partial Fulfillment of the Requirements for\\
  the Degree of Master of Philosophy \\
  in \programname \\
  \vfill \vfill
  \thesisdate, Hong Kong
  \vfill
\end{center}

\vfill

%% file: 2_authorization.tex
\null\skip0.2in
\begin{center}
{\bf \Large \underline{Authorization}}
\end{center}
\vspace{12mm}

I hereby declare that I am the sole author of this thesis.

\vspace{10mm}

I authorize the Hong Kong University of Science and Technology to lend this
thesis to other institutions or individuals for the purpose of scholarly research.

\vspace{10mm}

I further authorize the Hong Kong University of Science and Technology to
reproduce the thesis by photocopying or by other means, in total or in part, at the
request of other institutions or individuals for the purpose of scholarly research.

\vspace{30mm}

\begin{center}
\underline{~~~~~~~~~~~~~~~~~~~~~~~~~~~~~~~~~~~~~~~~~~~~~~~~~~~~~~~~~~~~~~~~~~~~~~}\\
~~~~\thesisauthor \\
~~~~\signdate

\end{center}

%% file: 3_signature.tex
\begin{center}
{\Large \thesistitle}\\
\vspace{5mm}
by\\
\vspace{3mm}
\thesisauthor\\
\vspace{5mm}
This is to certify that I have examined the above MPhil thesis\\
and have found that it is complete and satisfactory in all respects,\\
and that any and all revisions required by\\
the thesis examination committee have been made.
\end{center}

\vspace{15mm}

\begin{center}
\underline{~~~~~~~~~~~~~~~~~~~~~~~~~~~~~~~~~~~~~~~~~~~~~~~~~~~~~~~~~~~~~~~~~~~~~~~~~~~ }\\
\supervisorinfo
\end{center}

\vspace{15mm}
\begin{center}
\underline{~~~~~~~~~~~~~~~~~~~~~~~~~~~~~~~~~~~~~~~~~~~~~~~~~~~~~~~~~~~~~~~~~~~~~~~~~~~ }\\
\depheadinfo
\end{center}

\vspace{5mm}
\begin{center}
\departmentname\\
\vspace{5mm}
\signdate
\end{center}

%% file: 4_dedicate.tex
\thispagestyle{empty}
\null\vskip0.5in
\begin{center}

  \vspace{20mm}

  \begin{LARGE}
    \dedicate
  \end{LARGE}

  \vspace{4mm}

\end{center}

\vfill

%% file: 5_acknowledgement.tex
\centerline{{\bf \Large Acknowledgments}} \vspace{5mm} \noindent

First of all, I am genuinely thankful to my MPhil supervisor, Prof. Amir Goharshady. Research has been my primary goal for many years. However, my research journey was full of obstacles and discouragement, such that I was on the edge of giving up. Very luckily, I met Amir and he helped me change my life experience. Amir introduced me to theoretical computer science and research in blockchain. I was truly interested in these topics but had never got a chance to explore them before my MPhil. Amir's support in my research is beyond my imagination. He designed a great learning plan for me to prepare for research. He helped me and trained me through weekly reading groups and one-to-one meetings. Amir introduced me to an interesting research project on generating random numbers for blockchain, which leads to conference publications and also my MPhil thesis. Besides extensive technical discussions, Amir constantly motivated me when I was feeling down or stuck. Amir is also my role model now. I will continue my journey with Amir for my incoming PhD years. I am looking forward to learning more from Amir and transforming myself into an excellent PhD. 

I am also grateful to my lovely and friendly colleagues in Amir's research group, the ALPACAs. Specifically, I am thankful to Soroush Farokhnia and Hitarth Singh, who sat together with me for research projects and supported me tremendously. I am thankful to Ahmed Zaher, a great friend that took care of me many times. I also want to thank Togzhan Barakbayeva, Giovanna Kobus Conrado, Kerim Kochekov, Sergei Novozhilov, Pavel Hudec, Chun Kit Lam, Pingjiang Li, and other members of ALPACAS for great accompaniment in both academic meetings and social events. 

I am thankful to HKUST and Hong Kong. I enjoy my MPhil life, in terms of both academic achievement and social well-being. 

I also want to thank Weiyu Chen, Hansi Yang, Zhefan Rao, Ning Lu, Yunhao Gou, Nai Chit Fung, and many more best friends at HKUST. With these friends, my HKUST life is happier than ever before. I am thankful to Siqi Xie, my lovely girlfriend who supported me warmly and provided nice suggestions on my thesis defense presentation. Moreover, I want to thank Chang Lu and Zhengquan Cai, my parents, who brought me into this world and cared about me throughout my life. 

Finally, I am thankful to my thesis committee members for reviewing my thesis and offering valuable advice. 

%% file: abstract.tex
\begin{center}
{\Large \thesistitle}\\
\vspace{20mm}
by \thesisauthor\\
\departmentname\\
The Hong Kong University of Science and Technology
\end{center}
\vspace{8mm}
\begin{center}
Abstract
\end{center}
\par\noindent Many protocols in distributed computing rely on a source of randomness, usually called a random beacon, both for their applicability and security. This is especially true for proof-of-stake blockchain protocols in which the next miner or set of miners have to be chosen randomly and each party's likelihood to be selected is in proportion to their stake in the cryptocurrency. 

\par Current random beacons used in proof-of-stake protocols, such as Ouroboros and Algorand, have two fundamental limitations: Either (i)~they rely on pseudorandomness, e.g.~assuming that the output of a hash function is uniform, which is a widely-used but unproven assumption, or (ii)~they generate their randomness using a distributed protocol in which several participants are required to submit random numbers which are then used in the generation of a final random result. However, in this case, there is no guarantee that the numbers provided by the parties are uniformly random and there is no incentive for the parties to honestly generate uniform randomness. Most random beacons have both limitations. 

\par In this thesis, we provide a protocol for distributed generation of randomness. Our protocol does not rely on pseudorandomness at all. Similar to some of the previous approaches, it uses random inputs by different participants to generate a final random result. However, the crucial difference is that we provide a game-theoretic guarantee showing that it is in everyone's best interest to submit uniform random numbers. Hence, our approach is the first to incentivize honest behavior instead of just assuming it. Moreover, the approach is trustless and generates unbiased random numbers. It is also tamper-proof and no party can change the output or affect its distribution. Finally, it is designed with modularity in mind and can be easily plugged into existing distributed protocols such as proof-of-stake blockchains.

%% file: chapter1.tex
\chapter{Introduction}

\paragraph{Proof of Work} Bitcoin, the first blockchain protocol, was proposed by Satoshi Nakamoto to achieve consensus in a decentralized peer-to-peer electronic payment system \citep{nakamoto:bitcoin}. In Bitcoin and many other cryptocurrencies, the miners are selected by a proof-of-work (PoW) mechanism to add blocks of transactions to the public ledger~\cite{DBLP:conf/blockchain2/MeybodiGHS22}, i.e.~they have to compete in solving a mathematical puzzle and each miner's chance of adding the next block is proportional to their computational (hash) power. Security guarantees are then proven with the assumption that more than half of computational power is in the hands of honest miners. Proof of work is known to be highly energy-inefficient~\citep{bitcoin-energy,DBLP:conf/sac/ChatterjeeGP19} and also prone to centralization due to large mining pools \citep{weinberg:bitcoin-oligopoly}. Currently, the three largest mining pools have more than half of the entire Bitcoin mining power. 

\paragraph{Proof of Stake~\citep{kiayias:ouroboros}} Proof of Stake (PoS) is the main alternative consensus mechanism proposed to replace PoW in blockchain protocols. In a PoS protocol, miners are chosen randomly and each miner's chance of being allowed to add the next block is normally proportional to their stake in the currency. Hence, instead of relying on the assumption that a majority of the computational power on the network is owned by honest participants, the security claims of proof-of-stake protocols rely on the assumption that a majority, or a high percentage, of the stake is owned by honest participants. Despite their differences, all proof-of-stake protocols require a random beacon to randomly select the next miners in an unpredictable manner.

\paragraph{Distributed Randomness} A random beacon is an ideal oracle used in a distributed protocol, e.g.~a proof-of-stake blockchain, that emits a fresh random number in predetermined intervals. Designing random beacons is an active research topic in the context of distributed and decentralized computation \citep{bryan:scalable-bias-resistant-distributed-randomness,edgar:hydrand,gang-randchain,yuri:no-dealer-rng,icbc}. The desired properties of a random beacon are as follows:
\begin{itemize}
	\item \emph{Bias-resistance:} The output should always be sampled according to a fixed underlying distribution $\delta$, which is usually the uniform distribution. No party should be able to bias the output or change the distribution $\delta$. 
	\item \emph{Unpredictability:} No party should be able to predict the output before it is publicized. Moreover, no party should even be able to have any extra information about the output, other than the fact that it will be sampled from $\delta$. 
	\item \emph{Availability:} Each execution of the beacon must successfully terminate and produce a random value.
	\item \emph{Verifiability:} Each execution of the beacon should provide a ``proof'' such that any third party, even if not involved in the random beacon, is able to verify both the output and the fact that the random beacon executed successfully.
\end{itemize}

\paragraph{Reliable Participants} Almost all distributed randomness protocols have several participants and create a random output based on random numbers submitted by participants of the protocol. Usually, the final value is simply defined by the modular sum of all input values by participants modulo some large number $m,$ i.e.~$s := \sum_{i=1}^n s_i \pmod m.$ If the protocol generates only a single random bit, then $m=2$ and the modular sum is equivalent to the \texttt{xor} operation. Using the summation formula above, if the input values of different participants are chosen independently and if at least one of the participants submits a uniform random value in the range $[0, m-1]$, then the final output is also a uniform random value. We call such a participant \emph{reliable}. Note that it is enough to have only one reliable participant for the final output to have the uniform distribution when the values are submitted independently. Therefore, the distributed randomness protocols typically assume that at least one of the participants is reliable. In contrast, we distinguish between \emph{reliable} and \emph{honest} participants.

\paragraph{Honest Participants} An honest participant is a participant who correctly follows the protocol, e.g.~submits their random number $s_i$ in time. Distributed randomness protocols often assume and require that a large proportion of participants are honest and obey the communication rules to complete and produce final values. For example, PBFT achieves Byzantine agreement in a partially-synchronous network by requiring that more than two thirds of all participants be honest~\citep{castra:pbft}. 

\paragraph{Commitment Schemes} Using the formula above for random number generation, since the participants cannot broadcast their values in a distributed network in a perfectly simultaneous way, the last participant has an advantage and can dominate the final output. The classical cryptographic primitive used to avoid such a scenario is a commitment scheme. A commitment scheme runs in two phases: a commit phase and a reveal phase. In the commit phase, instead of broadcasting the value $s_i$ directly, each party $p_i$ broadcasts $h(s_i, r_i)$, where $h$ is a cryptographic hash function and $r_i$ is a randomly chosen nonce. In the reveal phase, each party broadcasts the values of $s_i$ and $r_i$ and everyone on the network can verify that the broadcast values have the right hash and thus the party has not changed their choice $s_i$ since the commit phase. However, a commitment scheme does not ensure availability, since malicious parties might only commit but not reveal their values. 

\paragraph{PVSS} Publicly verifiable secret sharing (PVSS) is a powerful cryptographic tool to ensure the revelation of values $s_i$ even if a number of malicious parties stop participating in the reveal phase of a commitment scheme~\citep{berry:simple-pvss}. PVSS adds a protection layer to traditional secret sharing schemes in the presence of malicious participants. In a PVSS scheme, a dealer is required to provide a non-interactive zero-knowledge proof (NIZK) along with encrypted secret shares $E_i(s_i)$ to guarantee the validity of secret shares. During the reconstruction phase, a participant sends their secret share to other participants along with an NIZK proof to guarantee the correctness of secret share. The NIZK proofs can be verified by any party, including third parties who are not taking part in the PVSS scheme. 

\paragraph{RANDAO~\citep{randao}} RANDAO is a family of smart contracts that produce random numbers. Anyone can participate and submit a random value to contribute to the output. RANDAO uses a commitment scheme. Compared to general distributed randomness protocols based on distributed networks, RANDAO's smart contracts run on a blockchain with consensus and directly interact with the underlying cryptocurrency. Therefore, RANDAO naturally enjoys the decentralized consensus provided by the blockchain protocol. Besides, economic incentives can be designed to promote honesty. Cheaters who violate the rules are punished economically, e.g.~by having their deposits confiscated. On the other hand, honest participants are rewarded by the income generated from providing the random number generation service to external contracts. However, there is no way to ensure bias-resistance and availability. A malicious party might choose not to reveal their value $s_i$ as it might be beneficial to them to bias the output. So, if a party does not reveal values, the whole random number generation process should be repeated, but even this biases the output as a malicious party can choose not to reveal only when the final result is not to their benefit in an external smart contract. Finally, RANDAO does not incentivize reliability and assumes that a reliable party exists, without arguing why.

\paragraph{EVR~\cite{yakira2020economically}} Another related approach is Economically Viable Randomness (EVR), which also uses a commit-reveal scheme to generate randomness. It designs a punishment scheme to discourage deviating from the protocol. However, similar to RANDAO, the incentives of EVR only care about whether the values are revealed faithfully. They do not differentiate between a reliable participant who submits a fresh uniformly-sampled random number and an unreliable honest participant who submits a constant number each time while following the rest of the protocol.

\paragraph{VDFs} Verifiable delay functions \citep{boneh:vdf} can be used to ensure bias-resistance in distributed randomness protocols. A VDF is a function whose evaluation takes at least some predetermined number of sequential steps, even with many parallel processors. Once the evaluation is complete, it can provide a publicly verifiable proof for the evaluation result, which can also be checked by any third party efficiently. 

\paragraph{VRFs} Verifiable random functions \citep{micali:vrf,yevgeniy:short-vrf} are widely used in PoS blockchain protocols \citep{micali:algorand,bernardo:ouroboros-praos}. A party can run a VRF locally, producing a pseudo-random output value based on their secret key and random seed. The VRF also outputs a proof of the output that can be verified by anyone with access to the party's public key and random seed. With the use of VRFs, malicious parties cannot predict who the future miners are before the miners announce their identities themselves. 

\paragraph{Algorand} Algorand \citep{micali:algorand} is a proof-of-stake blockchain protocol based on Byzantine agreement. The random seed for its VRF is based on the VRF of the previous round. While this guarantees most of the desired properties, a major drawback of this randomness beacon is that the generated numbers are not guaranteed to be uniform. 

\paragraph{Ouroboros and Ouroboros Praos} Ouroboros \citep{kiayias:ouroboros} was the first provably secure proof-of-stake blockchain protocol. It uses a publicly verifiable secret sharing scheme to generate a fresh random seed for each epoch. However, in this scheme the participants have no incentive to submit a uniform random value. In other words, there is no incentive to be reliable, but just to be honest. Ouroboros Praos \citep{bernardo:ouroboros-praos} improves over Ouroboros to be provably secure under a semi-synchronous setting. The random seed of Ouroboros Praos is updated every epoch by applying a random oracle hash function to a concatenation of VRF outputs in the previous epoch. Similar to Algorand, the random numbers are not guaranteed to be uniformly random, despite the fact that they are assumed to be uniform in the security analysis. 

\paragraph{Our Contributions} Our main contributions are as follows:
\begin{itemize}
	\item First, we design a novel game-theoretic approach for randomness generation. We call this an RIG (Random Integer Generation) game. RIG efficiently produces a uniform random integer from an arbitrarily large interval. Moreover, we show that the only equilibrium in an RIG is for all participants to choose their $s_i$ uniformly at random. In other words, our RIG ensures that the participants are incentivized not only to be honest, but also to be reliable. This will alleviate the problems with the previous approaches and ensure that all desired properties of distributed randomness are attained.
	
	\item We show that our RIG approach can be plugged into common randomness generation protocols with ease. In Chapter~\ref{section:rig-implementation}, we design protocols to implement RIG as a random beacon on general proof-of-stake blockchains. We describe RIG protocols based on commitment schemes and VDFs in Chapter~\ref{subsection:rig-commitment} and RIG protocols based on PVSS in Chapter~\ref{subsection:rig-pvss}. 
	
	\item In Chapter \ref{section:rig-in-blockchains}, we discuss how RIG can be deployed with minor changes in particular proof-of-stake protocols. We cover Algorand~\citep{micali:algorand} and Ouroboros Praos~\citep{bernardo:ouroboros-praos}. 
	
\end{itemize}

Our protocols are the first to incentivize participants to be reliable and submit uniform random numbers. In comparison, previous distributed randomness protocols using commitment schemes and PVSS assume that there is at least one reliable participant without incentivizing reliability. In other words, they only reward honesty but assume both honesty and reliability. The reliability assumption is unfounded.

Several other randomness protocols, including Algorand and Ouroboros Praos, do not depend on random inputs from participants at all, but instead use real-time data on blockchains and cryptographic hash functions to generate pseudo-random numbers. This pseudo-randomness is not guaranteed to be uniform, even though it is standard to assume its uniformity in security analyses. Hence, there is no guarantee that miners get elected with probabilities proportional to their stake. 

%% file: chapter2.tex
\chapter{Preliminaries and Related Works}

\section{Games and Equilibria}

\paragraph{Probability Distributions} Given a finite set $X=\{x_1, \dots, x_m\}$, a probability distribution on $X$ is a function $\delta:X\to [0, 1]$ such that $\delta(x_1) +\cdots + \delta(x_m) = 1$. We denote the set of all probability distributions on $X$ by $\Delta(X)$. 

\paragraph{One-shot Games \citep{algorithmic-game-theory-roughgarden}} A $\textit{one-shot game}$ with $n$ players is a tuple $G=(S_1, S_2, \dots, S_n,$ $u_1, u_2, \dots, u_n)$ where:
\begin{itemize}
\item Each $S_i$ is a finite set of \textit{pure strategies} for players $i$ and $S=S_1\times S_2\times\cdots \times S_n$ is the set of all \textit{outcomes}; and
\item Each $u_i$ is a \textit{utility function} of the form $u_i:S\to \mathbb{R}$.  
\end{itemize} 
\par In a \emph{play}, each player $i$ chooses one strategy $s_i\in S_i$. The choices are simultaneous and independent. Then each player $i$ is paid a $utility$ of $u_i(s_1, s_2, \dots, s_n)$ units. 

\paragraph{Mixed Strategies \citep{algorithmic-game-theory-roughgarden}} A \textit{mixed strategy} $\sigma_i \in \Delta(S_i)$ for player $i$ is a probability distribution over $S_i$, that characterizes the probability of playing each pure strategy in $S_i$. A \textit{mixed strategy profile} is a tuple $\sigma=(\sigma_1, \sigma_2, \dots, \sigma_n)$ consisting of one mixed strategy for each player. The \textit{expected utility} $u_i(\sigma)$ of player $i$ in a mixed strategy profile $\sigma$ is defined as $u_i(\sigma) = \mathbb{E}_{s_i \sim \sigma_i}[u_i(s_1, s_2, \dots, s_n)]$. 
Intuitively, in a mixed strategy, the player is not committing to a single pure strategy, but only to the probability of playing each pure strategy.

\paragraph{Nash Equilibria \citep{non-cooperative-games-nash}} A \textit{Nash Equilibrium} of a game $G$ is a mixed strategy profile $\sigma$, such that no player has an incentive to change their mixed strategy $\sigma_i$, assuming they are aware of the mixed strategies played by all the other players. Let $\sigma_{-i}$ be a tuple consisting of all the mixed strategies in $\sigma$ except $\sigma_i$. Formally, $\sigma$ is a Nash equilibrium if and only if for all $\tilde{\sigma}_i\in \Delta(S_i)$ we have $u_i(\sigma)\ge u_i(\tilde{\sigma}_i, \sigma_{-i})$. A seminal result by Nash is that every finite game $G$ has a Nash equilibrium \citep{non-cooperative-games-nash}. 
Nash equilibria are the central concept of stability and self-enforceability for non-cooperative games \citep{algorithmic-game-theory-roughgarden}, especially in the context of game-theoretic analysis of blockchain protocols~\cite{DBLP:conf/blockchain2/Goharshady21,DBLP:conf/concur/ChatterjeeGIV18,DBLP:conf/esop/ChatterjeeGV18,icbc2,sac23}, in which each player maximizes their own utility, i.e.~when a game is in a Nash equilibrium, no party has an incentive to change their strategy and hence the game remains in the Nash equilibrium.

In distributed randomness generation, especially for proof-of-stake protocols, we aim to have a committee that plays a game whose output is our random number. Since the players/parties are pseudonymous on a blockchain network and only participate using their public keys, we might have multiple accounts in our committee that are actually controlled by the same person or are in an alliance. Therefore, we need a stronger concept of equilibrium that does not assume a lack of cooperation between any pair of players. Thus, we rely on strong and alliance-resistant equilibria as defined below.

\paragraph{Strong Nash Equilibria \citep{acceptable-n-games,coalition-proof-nash-equilibria}} A \textit{strong Nash equilibrium} is a mixed strategy profile in which no group of players have a way to cooperate and change their mixed strategies such that the utility of every member of the group is increased. Formally, $\sigma$ is a strong Nash equilibrium if for any non-empty set $P$ of players and any strategy profile $\tilde{\sigma}_P$ over $P$, there exists a player $p\in P$ such that $u_p(\sigma)\ge u_p(\tilde{\sigma}_P, \sigma_{-P})$. 
In strong equilibria, the assumption is that the players cannot share or transfer their utilities, so a player agrees to a change of strategies in the alliance $P$ if and only if their own utility is strictly increased. However, in cases where the utitlity is fungible, i.e.~if the players can share and redistribute utilities, or if they are indeed controlled by the same person, then a group is willing to defect as long as their \textit{total} utility increases, which leads to an even stronger notion of equilibrium: 

\paragraph{Alliance-resistant Nash Equilibria~\citep{krish:probabilistic-smart-contracts}} An \textit{alliance-resistant Nash equilibrium} is a mixed strategy profile $\sigma$ such that for any non-empty set $P$ of players and any strategy profile $\tilde{\sigma}_P$, it holds that $u_P(\sigma)\ge u_P(\tilde{\sigma}_P, \sigma_{-P})$, where $u_P$ is the sum of utilities of all members of $P$. 

In our setting, especially in PoS blockchain protocols, alliance-resistant equilibria are the suitable notion to justify stability and self-enforceability, because a person with a large stake is likely to control multiple players in the randomly-selected committee and only care about the overall revenue. 

\section{Commitment Schemes} 
Using commitment schemes is a common paradigm to address the vulnerability of malicious domination by the last participant. A commitment scheme is a cryptographic protocol that allows a sender to commit a chosen value while keeping it hidden to the receiver, and the receiver has the ability to access the committed value later with the help of the sender. A commitment scheme has two phases: (i)~In the \emph{commit} phase, the sender holds a message $s$ and chooses a random string $r\in \{0,1\}^\kappa$, where $\kappa$ is a security parameter. The sender then encodes them to $c$ and sends $c$ to the receiver; (ii)~In the \emph{reveal} phase, the sender sends some hint $w$ to the receiver. Using this $w,$ the receiver can open the commitment $c$ to get $x$. A commitment scheme should satisfy the following two security properties:

\begin{itemize}
     \item \textbf{Hiding.} Receiving a commitment $c$ should leak no information about message $s$. Formally, for $\forall s_0$, $s_1$, let $p_0\sim \{(r,c)|r\overset{{\scriptscriptstyle\$}}\gets \{0,1\}^\kappa, c\overset{{\scriptscriptstyle\$}}\gets \mathbf{Commit}(s_0, r) \}$,  $p_1\sim \{(r,c)|r\overset{{\scriptscriptstyle\$}}\gets \{0,1\}^\kappa, c\overset{{\scriptscriptstyle\$}}\gets \mathbf{Commit}(s_1, r) \}$. Then the distribution of $p_0$ and $p_1$ should be computationally indistinguishable. 
     \item \textbf{Binding.} Values other than $x$ should not be encoded into the same commitment $c$. Formally, for all non-uniform probabilitistic polynomial time algorithms that output $s_0$, $s_1$ and $r_0$, $r_1$, the probability that $s_0\neq s_1$ and $\mathbf{Commit}(s_0, r_0)=\mathbf{Commit}(s_1, r_1)$ is a negligible function in $\kappa$, the length of $r_0$ and $r_1$. 
\end{itemize}

A simple instantiation of a commitment scheme is through a cryptographic hash function $hash$, such as the SHA256, which generates a $256$-bit string for an input string of arbitrary length. The value $s$ to commit might be a single bit or an element from an input space $\mathcal{S}\subseteq \{0,1\}^\star$. The sender chooses a random string $r\in\{0,1\}^\kappa$ and computes $c = hash(s||r)$, where $||$ refers to bit concatenation. If an adversary with bounded computational resources only knows the commitment value $c$, she knows very little information about $s$. Even if $s$ is a random bit, which is either $0$ or $1$, the sender can hide $s$ by choosing a large enough $\kappa$ for the random string $r$. In the reveal phase, the sender simply sends both $s$ and $r$ to the receiver. The receiver can compute $hash(s||r)$ using the received $s$ and $r$ and compare it with the commitment. It is computationally infeasible for the sender to come up with another pair of $s^\prime$ and $r^\prime$ such that $hash(s^\prime||r^\prime)=hash(s||r)$, guaranteed by the cryptographic hash function. Therefore we can implement a commitment scheme using hash functions. 

With a commitment scheme, the participants of RNG can commit their values in the commit phase and announce the values in the reveal phase after every participant sends their commitment. The last participant does not know the values of other participants when she chooses her value in the commit phase, hence cannot arbitrarily tamper the output. However, she might maliciously choose not to reveal her value in the reveal phase and it is computationally infeasible for other participants to open her commitment. If the protocol computes the output without her value, then she has still succeeded in tampering with the output. In the case of a single random bit, she can even dominate the output value. Even if we design a punishment scheme by claiming a pre-specified amount of deposit for such malicious behavior, it is not effective if the output has significant economic consequences and the benefits from the tampering potentially outweigh the confiscated deposits. 

\section{Publicly-Verifiable Secret Sharing (PVSS)} \label{sec:pvss}

We follow~\citep{berry:simple-pvss} in our description of PVSS.  In a PVSS scheme, a dealer $D$ wants to share a secret $s$ with a group of $n$ participants $P_1, P_2, \dots, P_n$. The goal is to have a $(t, n)$-threshold scheme, i.e.~any subset of $t$ participants can collaborate to recover the secret $s$, while any smaller subset of participants cannot recover the secret or obtain any information about it. Moreover, anyone on the network, even those who are not participating, should be able to verify that the dealer is acting honestly and following the protocol.

\paragraph{Initialization} We assume that a multiplicative group $\mathbb{Z}_q^*$ and two generators $g, G$ of this group are selected using an appropriate public procedure. Here, $q$ is a large prime number and all calculations are done modulo $q$. Each participant $P_i$ generates a non-zero private key $x_i\in \mathbb{Z}^*_{q}$ and announces $y_i = G^{x_i}$ as their public key. 
Suppose the secret to be shared is $s$, the dealer first chooses a random number $r$ and publishes $U= s + h(G^r)$, where $h$ is a pre-selected cryptographic hash function. The dealer then runs the main protocol below to distribute the shares that can reveal $G^r$. 
The main protocol consists of two steps: (1)~distribution, and (2)~reconstruction, each of which has two substeps.

\paragraph{Distribution} This consists of the following substeps:
\begin{itemize}

\item \emph{Distribution of the shares}. The dealer picks a random polynomial $p$ of degree at most $t-1$ with coefficients in $\mathbb{Z}_q$ of the form
$
 \textstyle	p(x) = \sum_{j=0}^{t-1}  \alpha_j \cdot x^j.
$
Here, we have $\alpha_0 = G^r$, i.e.~the number $r$ is encoded in the first coefficient of the polynomial and every other $\alpha_j$ is a random number from $\mathbb Z_q$. The dealer then publishes the following:
\begin{itemize}
	\item \emph{Commitment:} $C_j = g^{\alpha_j}$, for $0\le j < t$. This ensures that the dealer is committing to the polynomial and cannot change it later. 
	\item \emph{Encrypted shares:} For each player $P_i$, the dealer computes and publishes $Y_i = y_i^{p(i)}$, for $1\le i \le n$. Intuitively, the dealer is taking the value $p(i)$ of the polynomial $p$ at point $i$ and encrypting it using $y_i$ so that only the $i$-th player can decrypt it. This encrypted value is then published.
	\item Proof of correctness: The dealer provides a non-interactive zero-knowledge proof ensuring that the encrypted shares above are valid. See~\citep{berry:simple-pvss} for details. 
\end{itemize}

\item \emph{Verification of the shares.} Anyone on the network, be it a player $P_i$ or a non-participant third-party, can verify the proof and encrypted shares provided by the dealer to ensure that the dealer is acting honestly, i.e.~following the protocol above, and is not giving out invalid shares.
\end{itemize}

\paragraph{Reconstruction} This step consists of:

\begin{itemize}
\item\emph{Decryption of the shares.} Each party $P_i$ knows $Y_i = y_i^{p(i)}$ and their secret key $x_i.$ Recall that $y_i = G^{x_i}.$ Hence, the $i$-th party can compute $Y_i^{1/x_i} = y_i^{{p(i)}/{x_i}} = G^{p(i)}.$ They publish $G^{p_i}$ along with a non-interactive zero-knowledge proof of its correctness.  
\item \emph{Pooling the shares.} Any $t$ participants $P_{i_1}, P_{i_2}, \dots, P_{i_t}$ can compute the $G^r$ by Lagrange interpolation. More specifically, they know $t$ points $(i_j, p(i_j))$ of the polynomial $p$ that is of degree $t-1$. So, they can find the unique polynomial that goes through these points. Note that after all the shares are decrypted, anyone on the network can use $t$ of the shares to compute the polynomial $p$ and then $G^r$ is simply $p(0).$ However, before the decryption of the shares in the reconstruction step, finding $G^r$ requires the collaboration of at least $t$ participants and no set of $t-1$ participants can obtain any information about $G^r$. Finally, knowing $G^r$ and $U$, it is easy to find the secret $s,$ i.e.~$s = U - h(G^r).$
\end{itemize}
PVSS schemes can be used to generate random numbers. To do so, we use a separate PVSS scheme for each participant $P_i$. All $n$ PVSS schemes run in parallel. In the $i$-th scheme, $P_i$ is the dealer and everyone else is a normal participant. $P_i$ first chooses a random number $s_i$ and then performs the protocol above as the dealer. At the end of the process, all the $s_i$'s are revealed by pooling the shares and we can use $s = \sum_{i=1}^n s_i$ as our random number. The upside is that no party can avoid revealing their $s_i$ and hence the protocol satisfies availability. The downside is that every set of $t$ parties can unmask everyone else's choices and hence bias the result. Therefore, in random number generation using PVSS we have to assume that there are at most $t-1$ dishonest participants. Additionally, since each party has to send shares to every other party, the protocol will have a quadratic communication complexity.

\section{Verifiable Delay Functions (VDFs)}
We follow \citep{boneh:vdf} in our treatment of verifiable delay functions. A verifiable delay function (VDF) is a tuple $V=(\mathbf{Setup}, \mathbf{Eval}, \mathbf{Verify})$ parameterized by a security parameter $\lambda$ and a desired puzzle difficulty $t$. Suppose our input space is $\mathcal{X}$ and our output space is $\mathcal{Y}$. $V$ is a triplet of algorithms as follows: 
\begin{itemize}
	\item $\mathbf{Setup(\lambda, t)}\to (ek, vk)$. This function generates an evaluation key $ek$ and a verification key $vk$ in polynomial time with respect to $\lambda$.  
	\item $\mathbf{Eval}(ek, x)\to (y, \pi)$ takes an input $x\in\mathcal{X}$ and produces an output $y\in\mathcal{Y}$ and a proof $\pi$. $\mathbf{Eval}$ may use randomness in computing the proof $\pi$ but not in the computation of $y$. It must run in parallel time $t$ with $poly(\log(t), \lambda)$ processors. 
	\item $\mathbf{Verify}(vk, x, y, \pi)\to \{\text{Yes, No}\}$ is a deterministic algorithm that verifies the correctness of evaluation in sequential time $poly(\log(t), \lambda)$. 
\end{itemize}
See \citep{boneh:vdf} for more details and desired properties. 
Intuitively, anyone can evaluate the VDF using the evaluation key. However, this takes a long time, i.e.~at least $t$ steps, even when using parallelization. Even if a malicious participant has strong parallel computational power, they cannot evaluate the VDF significantly faster than an ordinary participant that owns only a single processor. However, after the evaluation is done, verifying the result is easy and much faster and anyone can do the verification using the verification key $vk$.

The use-case of verifiable delay functions in random number generation is to again defend against dishonest participants who do not reveal their choice in a commitment scheme. We can require every participant to provide a VDF whose evaluation is their choice $s_i$. Then, even if the participant is dishonest and does not reveal their own choice, other participants can evaluate the VDF and obtain the $s_i$, hence ensuring availability for our random number generation protocol. However, evaluation takes a long time, and hence the choice will not be revealed while in the commit phase.

Note that both PVSS and VDF methods above can be used to ensure availability and defend against dishonest parties who do not reveal their choices. However, they do not incentivize the parties to be reliable and choose their $s_i$ uniformly at random. This is our main contribution in the next chapter.

\section{Lottery Protocols}
Lottery is an important problem in distributed computation~\cite{DBLP:conf/podc/Broder85} and also a financial activity that has a large market capitalization~\cite{lottery-market}. One winner or a set of winners are randomly selected from the participants to win an award. It is desirable to select the winner randomly without trusting a lottery company or other third parties. Therefore, online lottery schemes typically require decentralized randomness. Our RIG protocol can be applied to lottery schemes to achieve guaranteed random selection. Furthermore, depending on the application of RIG, our protocol can also benefit from techniques proposed in the community of lottery schemes. For example, besides the \texttt{xor} or modular addition methods to aggregate the randomness contribution of each participant, other aggregation methods have been studied in the context of lottery schemes~\cite{DBLP:journals/ijcomsys/LiuLCCJ14}. Privacy-preserving properties are also studied, for example using blind signatures~\cite{DBLP:journals/csi/LeeCI09}. We summarize how online lottery schemes use random number generators. 

\paragraph{Online Lottery Schemes} There are many online lottery schemes in the literature. Here is a non-exhaustive list of some of the most notable works in this direction:
\begin{itemize}
	
	\item Chow~\cite{DBLP:conf/iccsa/ChowHYC05} proposed an online lottery scheme using a hash chain to link the lottery tickets of participants. A verifiable random function (VRF) is applied to extract verifiable randomness from the hash chain and a verifiable delay function (VDF) is applied to avoid the malicious adversary of the last participant. Similar to commitment schemes with VDFs, the sequential evaluation time of VDFs brings efficiency concerns and even fairness concerns due to different computational environment. 
	
	\item Lee~\cite{DBLP:journals/csi/LeeCI09} designed a scheme that uses the Chinese Remainder Theorem and Blind Signatures in the lottery tickets. For the randomness to select the winner, Lee's scheme uses a pseudo-random number generator and seeds it with the sequential modular sum of random values submitted by all the participants. These random values are sent to the lottery dealer under encryption. In this scheme, the last participant can collude with the lottery dealer to tamper the random seed by decrypting other participants' random values.
	
	\item Liu~\cite{DBLP:journals/ijcomsys/LiuLCCJ14} improved this protocol by changing the random seed to a Lagrange interpolation result that depends on random values of all participants. The random values are committed in the first phase. Liu's scheme does not address the issue of maliciously hiding the random values, as in the commitment schemes. 
	
	\item Grumbach~\cite{DBLP:conf/dais/GrumbachR17} follows the paradigm of using delay functions to handle manipulation. The contribution is to use a Merkle tree structure to achieve fast probabilitistic verification in large-scale lottery systems, which is orthogonal to our contribution. 
	
	\item Xia~\cite{DBLP:journals/symmetry/XiaLHC19} proposed a lottery scheme using symmetric bivariate polynomials to share random secrets among different lottery centers. This distributed randomness is similar to that of PVSS schemes.
\end{itemize}

\paragraph{Lottery Schemes on Blockchain} Blockchain and smart contracts are well-suited for distributed protocols without trusted third-parties. Thus, there are many works on blockchain-based lotteries, as well:
\begin{itemize}
\item Inspired by the random number generation of RANDAO~\cite{randao}, Jia~\cite{DBLP:journals/corr/abs-1911-02392} implemented DeLottery, a decentralized lottery and gambling system based on smart contracts. A considerable amount of deposit is required to punish malicious behavior in the commit-reveal method of RANDAO. 

\item Jo~\cite{DBLP:journals/corr/abs-1912-00642} implemented BlockLot on Hyperledger/Fabric blockchain platform which provides transparent, fair and verifiable lottery services. It hashes a random source such as block hash, extracts the first four 32-bit integers and uses the sum to randomly select a winner. This method is prone to manipulation by malicious miners and does not guarantee to select a winner uniformly randomly. 

\item Li~\cite{DBLP:conf/desec/LiZ019} presented a lottery smart contract where the randomness to select winners is drawn from a polynomial interpolation. The polynomial is defined by points chosen by each participant and the points are sent to the lottery center in the encrypted form. The encryption scheme requires the lottery center to be a trusted third party and not collude with lottery participants.

\item Chen~\cite{DBLP:conf/icbc2/ChenHCW19} proposed a lottery DApp where the randomness is determined by a combination of game states, blockchain states and a random number generation committee. 

\item Pan~\cite{DBLP:conf/blockchain2/PanZLWS22} designed FPLotto that uses Lagrange interpolating polynomial and VRFs to ensure all participants fairly contribute to the randomness. Further it uses zero-knowledge proofs to hide lottery amounts and uses aggregation verification to accelerate verification. 
\end{itemize}

In summary, all lottery schemes on blockchain use random numbers, either through (i)~pseudo-random number generators or random oracles that use blockchain states as seeds, or (ii)~an external random number generation protocol. Therefore, they all inherit the limitations of existing random number generation methods, while some of them present solutions to achieve orthogonal properties such as privacy protection. Thus, our approach, though mainly motivated by PoS, can help bring provably-uniform game-theoretic randomness to lotteries, as well.

%% file: chapter3.tex
\chapter{Random Integer Generation}\label{section:rig}

In this chapter, we provide the main component of our approach, i.e.~a game to incentivize reliability in random number generation.

\section{Overview of RIG}\label{subsection:rig-overview}

\paragraph{RIG} Suppose that we have $n$ players and $n$ is even. A \textit{Random Integer Generation game} (RIG) with $n$ players and $m \ge 2$ strategies is a game $G$ in which: 
\begin{itemize}
\item For every player $i \in \{1, \ldots, n\}$, we have $m$ pure strategies $S_i = \{0, 1, \dots, m-1\}$;
\item We pair the players such that every even player is paired with the previous odd player and every odd player is paired with the next even player. In other words, $pair(2\cdot k) = 2 \cdot k - 1$ and $pair(2 \cdot k - 1) = 2 \cdot k.$
\item At an outcome $s = (s_1, s_2, \dots, s_{n})$ of the game, the payoff of player $i=2\cdot k-1$ is defined as $u_i(s)\coloneqq  f(s_{2k-1}, s_{2k}) $ and the payoff of player $j=2\cdot k$ is defined as $u_j(s)\coloneqq -u_i(s) = -f(s_{2k-1}, s_{2k})$, 
where 

$$
\textstyle f(s_{2k-1}, s_{2k}) \coloneqq
\begin{cases}
	1       & \quad \text{if } s_{2k-1}-s_{2k}\equiv 0  \pmod m\\
	-1  & \quad \text{if } s_{2k-1}-s_{2k}\equiv -1 \pmod m\\
	0 & \quad \text{otherwise}
\end{cases}
$$
\end{itemize}
Essentially, we assume that any adjacent pair of even player and odd player play a zero-sum one-shot game with each other. Their payoffs are independent of the other $n-2$ players. For each pair $(2 \cdot k - 1, 2 \cdot k)$ of players, this is a zero-sum matrix game with the following payoff matrix $A^{(m)}$ for player $2\cdot k -1$:

$$
A^{(m)} = 
\begin{pmatrix}
	1 & -1 & 0 & \cdots & 0 \\
	0 & 1 & -1 & \cdots & 0 \\
	0 & 0 & 1 & \cdots & 0 \\
	\vdots  & \vdots  & \ddots & \vdots  \\
	-1 & 0 & 0 & \cdots & 1
\end{pmatrix}
$$

\begin{figure}[H] 
	\centering
	\includegraphics[width=0.9\textwidth]{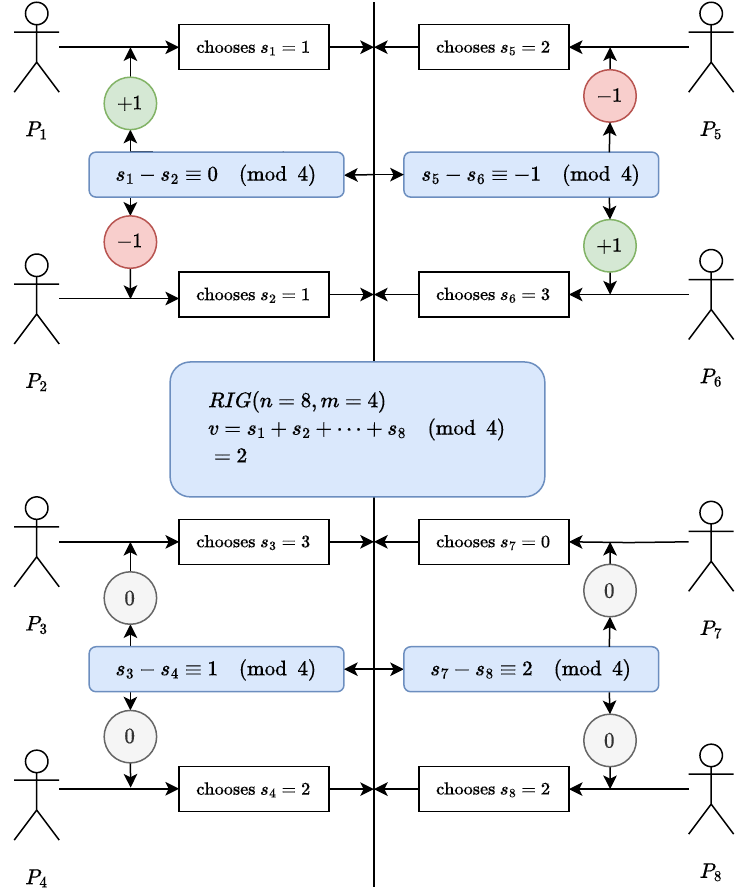}
	\caption{Example of RIG: $n=8$ players generate an integer in $\{0,1,2,3\} (m=4)$. }
	\label{fig:rig}
\end{figure}

An example RIG execution is presented in Figure~\ref{fig:rig}. In this example, there are $8$ players and the random output is in the set $\{0,1,2,3\}$. In other words, $n=8$ and $m=4$. The players are ordered as $P_1, P_2,\dots, P_8$. $P_1$ and $P_2$ form a pair. The other three pairs are $(P_3, P_4)$, $(P_5, P_6)$, $(P_7, P_8)$. For each pair, the payoffs are defined by the $4\times 4$ matrix $A^{(8)}$ and also depend on the values submitted by each participant. For example, $P_1$ chooses $s_1=1$ and $P_2$ chooses $s_2=1$, then $s_1-s_2\equiv 0\pmod 4$. According to the payoff matrix $A^{(8)}$, $P_1$ receives $1$ unit of reward from $P_2$. The output random integer of RIG is $$\sum_{i=1}^8s_i\pmod 4=2.$$ 

\section{Analysis of Alliance-Resistant Nash Equilibria}\label{subsection:rig-analysis}

We now show that our RIG has the desired game-theoretic properties. 

\begin{theorem} {(Alliance-Resistant Nash Equilibrium of an RIG.)}
Let $G$ be an RIG game with $n$ players and $m$ strategies, where $n$ is an even number and $m\ge 2$. Let $\bar{\sigma}$ be a mixed strategy profile defined by $\bar{\sigma}_i = (1/m, 1/m, \dots, 1/m)$ for all $i$, i.e.~the mixed strategy profile in which each player $i$ chooses a strategy in $S_i$ uniformly at random. Then, $\bar{\sigma}$ is the \textit{only} Nash equilibrium of $G$. Further, it is also alliance-resistant. 
\end{theorem}

\begin{proof} First, we prove that $\bar{\sigma}$ is an alliance-resistant Nash Equilibrium. Under the mixed strategy profile, the expected payoff of each one of the players is $0$. Let $G$ be a subset of players, then the overall utility of all players in $G$ is $\sum_{i\in G} u_i(\bar{\sigma}_{- G}, \sigma)$ if players in $G$ play another strategy profile $\sigma$. Each player $i$ is effectively playing against its adjacent player. If both player $i$ and player $pair(i)$ are in $G$, then $u_i(\bar{\sigma}_{- G}, \sigma) = - u_{pair(i)}(\bar{\sigma}_{- G}, \sigma)$. The utilities of these two players always sum up to zero, so that the overall utility of $G$ is not influenced by them. Similarly, if both player $i$ and player $pair(i)$ are not in $G$, they do not influence the overall utility of $G$ either. The only non-trivial part consists of those players in $G$ who play against players outside $G$. For each such player $i$, since the player $pair(i)$ plays mixed strategy $\bar{\sigma}_{pair(i)}$, the utility is $u_i = \sigma_i ^T \cdot  A^{(m)} \cdot \bar{\sigma}_{pair(i)} = \sigma_i^T\cdot (0, 0, \dots, 0)=0$. Therefore, the overall utility of $G$ is $0$ and changing the strategy has no benefit.

We now prove that $\bar{\sigma}$ is the unique Nash equilibrium of this game. Suppose there is another strategy profile $\tilde{\sigma}$ that is also a Nash equilibrium. Then for any player $i$, since it is effectively only playing with its adjacent player $j=pair(i)$, it follows that $(\tilde{\sigma}_i, \tilde{\sigma}_j)$ forms a Nash equilibrium for the zero-sum bimatrix game defined by $A^{(m)}$. 

Now consider the bimatrix game between player $i$ and player $j$. Let their utility at Nash equilibrium mixed strategy profile $(\tilde{\sigma}_i, \tilde{\sigma}_j)$ be $(\tilde{u}_i, \tilde{u}_j)$. Since it is a zero-sum matrix game, $\tilde{u}_i + \tilde{u}_j=0$. Without loss of generality, assume that $\tilde{u}_i \le \tilde{u}_j$, then $\tilde{u}_i \le 0$. By the definition of Nash equilibrium, player $i$ cannot increase its utility by changing its strategy $\tilde{\sigma}_i$ to any other strategy $\sigma_i$, while player $j$ keeps playing the same strategy $\tilde{\sigma}_j$. This indicates that every coordinate of the vector $A^{(m)}\cdot \tilde{\sigma}_j$ is no more than $\tilde{u}_i$, which is at most 0. Let $\tilde{\sigma}_j = (p_0, p_1, \dots, p_{m-1})$, then the utility of the player $i$ playing pure strategy $k$ is $\delta_k = p_k - p_{k+1\pmod m}\le\tilde{u}_i\le0$, for each $k$ in $\{0, 1, \dots, m-1\}$. However, $\sum_{k=0}^{m-1}\delta_k = \sum_{k=0}^{m-1} p_{k} - \sum_{k=0}^{m-1}p_k = 0$. So it must hold that $\tilde{\sigma}_{j}=(1/m, 1/m, \dots, 1/m)$ and  $\tilde{u}_i=\tilde{u}_j=0$. Since $\tilde{u}_j=0\le 0$, a similar analysis can show that $\tilde{\sigma}_i = (1/m, 1/m, \dots, 1/m)$. This proves that $\bar{\sigma}$ is the only Nash equilibrium for the Random Integer Generation game. 
\end{proof}

The theorem above shows that it is in every player's best interest to play uniformly at random, i.e.~choose each pure strategy in $S_i$ with probability exactly $1/m.$ Moreover, this equilibrium is self-enforcing even in the presence of alliances. Hence, we can plug this game into a distributed random number generation protocol and give participants rewards that are based on their payoffs in this game. This ensures that every participant is incentivized to provide a uniformly random $s_i$. As mentioned before, even if one participant is reliable and submits a uniformly random $s_i$, then the entire result $\textstyle v = \sum s_i$ of the random number generation protocol is guaranteed to be unbiased. Hence, instead of assuming that a reliable party exists, we incentivize every party to be reliable.

\section{Dense RIG Bimatrix Game}\label{subsection:dense-rig}
If the strategy size $m$ is large, which is the case when we aim to generate integers from a large range, then the matrix $A$ would be sparse. If the number of players is much smaller than $m$, then the probability that one party really receives a non-zero payoff is negligible. Therefore, it is desirable to design a matrix $B$ that is dense and also provides the same unique alliance-resistant equilibrium property as $A$. 

For any integer $f$ such that $1\le f \le m/2$ and $\texttt{gcd}(f,m)=1$, we can construct a new matrix $B^{(m,f)}$ with dense parameter $2f/m$ such that $B^{(m,f)}_{i, j}=g(j-i)$, where $g(\cdot)$ is defined as:
$$ \textstyle g(l) \coloneqq
\begin{cases}
1       & \quad \text{if } 0\le l \le f-1 \pmod m\\
-1  & \quad \text{if } m-f \le l \le m-1 \pmod m\\
0 & \quad \text{otherwise.}
\end{cases} 
$$

For $m=8$ and $f=3$, the new payoff matrix $B^{(8,3)}$ is the following: 
$$
B = 
\begin{pmatrix}
	1 & 1 & 1 & 0 & 0 & -1 & -1 & -1 \\
	-1 & 1 & 1 & 1 & 0 & 0 & -1 & -1 \\
	-1 & -1 & 1 & 1 & 1 & 0 & 0 & -1 \\
    -1 & -1 & -1 & 1 & 1 & 1 & 0 & 0 \\
    0 & -1 & -1 & -1 & 1 & 1 & 1 & 0 \\
    0 & 0 & -1 & -1 & -1 & 1 & 1 & 1 \\
    1 & 0 & 0 & -1 & -1 & -1 & 1 & 1 \\
    1 & 1 & 0 & 0 & -1 & -1 & -1 & 1
\end{pmatrix}
$$

When $f=1$, the matrix $B^{(m, 1)}$ is the same as $A^{(m)}$.  We can check that the mixed strategy profile $\bar{\sigma}$ under $B^{(m,f)}$ is also an alliance-resistant Nash equilibrium. To show that it is the only Nash equilibrium, we follow the analysis we did for $A^{(m)}$. Suppose there is another Nash equilibrium $(\tilde{\sigma}_i, \tilde{\sigma}_j)$ between two adjacent players $i$ and $j(=i+1)$ and $\tilde{u}_i\le0$. Let $\tilde{\sigma}_j$ be $(p_0, p_1, \dots, p_{m-1})$, then every element of $r=B \cdot \tilde{\sigma}_j $ is at most $\tilde{u}_i$, which is non-positive. However, $\sum_{k=0}^{m-1} r_k = \mathbf{1}^T \cdot r = \mathbf{1}^T \cdot B \cdot \tilde{\sigma}_j = \mathbf{0}^T \cdot \tilde{\sigma}_j = 0$, which requires $r=\mathbf{0}$ and $\tilde{u}_i=0$. By $r=\mathbf{0}$, we have
\begin{displaymath}
r_s = \sum_{0\le l\le f-1} p_{l+s}-\sum_{m-f \le l \le m-1} p_{l+s} = 0 
\end{displaymath}
for every $s\in\{0, 1, \dots, m-1\}$. With a slight misuse of notation, assume that $p_{m+t}=p_t$ for any integer $t$. 
If we subtract $r_{s+1}$ by $r_s$, we get 
\begin{displaymath} \textstyle
p_{s+f}-p_{s} = p_{s+m} - p_{s+m-f} = p_{s} - p_{s-f}
\end{displaymath}
Let $q(s)=p_{s}-p_{s-f}$, then $q(s+f) = q(s)$. We also have $q(s+m) = q(s)$. Since $\texttt{gcd}(f, m)=1$, the function $q(\cdot)$ is constant on integers, from which we can infer that $p_0=p_1=\cdots =p_{m-1}=1/m$. Therefore, $\bar{\sigma}$ is still the only Nash equilibrium.

\paragraph{Important Remark on Parallelization} Note that the parallelization of RIG loses the uniqueness property of Nash equilibrium, hence we cannot simply have an RIG game for $m=2,$ use it to generate a random bit, and parallelize it $k$ times to generate $k$ random bits. Instead, we must set $m = 2^k$ and have a single non-parallel RIG game. As an example, consider the simplified case of two players and $k$ bits. If each player only uniformly randomly set their first bit, and then copied the same bit to all other bits, this would also form a Nash equilibrium. However, this Nash equilibrium does not produce a uniform random output. Instead, the output is $0$ with $50\%$ probability and $2^k-1$ with $50\%$ probability. More generally, any $\sigma=(\sigma_1, \sigma_2)$ such that $\sigma_i(j\text{-th bit is }0)=1/2$ for all $1\le j\le m$ is a Nash equilibrium. The existence of these undesirable Nash equilibria breaks the guarantee of uniform random distribution over the final output value. Hence, parallelization should not be used and a game on $2^k$ strategies should be played in the first place. 

%% file: chapter4.tex
\chapter{Designing a Random Beacon Based on RIG} \label{section:rig-implementation}

In this Chapter, we discuss how to use a single execution of the Random Integer Generation game in a distributed random number generation beacon. The major challenge is to execute the game, in which the parties have to move simultaneously, in a decentralized environment. We propose two alternatives to implement the RIG game: (1)~using commitment schemes and verifiable delay functions, and (2)~using publicly-verifiable secret sharing. We assume that the set of players is already fixed. Usually, only a small subset of users (or blockchain miners) are selected from all the users in the system to join a single execution of the game (generation of a random number). The selection rule is determined by the specific application. The design and amount of reward/penalty and deposits are also subject to the specific application. We will address these adjustable configurations in Section 4.3. Naturally, we focus on the case where our protocol is used in conjunction with a blockchain protocol, including gossip protocols for public communication.  

\section{Random Beacon using Commitment Schemes and Verifiable Delay Functions}\label{subsection:rig-commitment}

As mentioned above, commitment schemes are already widely used for random number generation in distributed systems. As expected, our approach has two phases: commit and reveal. The execution starts with the commit phase, which lasts for $T_{commit}$ time slots. In a blockchain ecosystem, we can use the block number to keep track of time. After the commit phase ends, the execution enters the reveal phase, which lasts for $T_{reveal} $ time slots. The RIG game is executed as soon as the reveal phase completes. 

\paragraph{Commit Phase} In the commit phase, each participant $P_i$ broadcasts a signed commit message, in the form of $(ss\_id, h_i, proof_i)_i$, where $ss\_id$ is the session id of the execution and $h_i=hash(s_i | r_i)$ is the commitment. $s_i$ is the value the participant chooses and $r_i$ is a random nonce. $proof_i$ is a publicly-verifiable proof that $P_i$ is an eligible participant, applicable when only a specific subset of selected users are allowed to join the game. A commit message is valid if:
\begin{enumerate}[(1)]
	
	\item The message is properly signed by $P_i$;
	\item The message has a valid proof of participation;
	\item There is no other different valid commit message from $P_i$ in the network;
	\item The message is received during the commit phase; and
	\item Valid VDF parameters are provided in the message (discussed further below).
\end{enumerate}

\paragraph{Reveal Phase} In the reveal phase, each participant $P_i$ broadcasts a signed reveal message, in the form of $(ss\_id, s_i, r_i)_i$.  A reveal message is valid if
\begin{enumerate}[(1)]
\item The message is received during the reveal phase; 
\item $P_i$ has exactly one valid commit message; and 
\item $hash(s_i | r_i)$ matches the commitment $h_i$ of the participant $P_i$. 
\end{enumerate}

\paragraph{RIG} After the reveal phase completes, we can compute the payoffs of the RIG game. Assume the outcome of the game is $(s_1, \ldots, s_n)$, where $s_i \in \{0, 1, \dots, m-1\}$ is the strategy played by each player. We set $v := \sum s_i \pmod m$ and output it as the result of the random number generation protocol. 

\paragraph{Role of the VDF} The value of $r$ can be biased by malicious participants who might choose not to reveal their values/strategies. If a participant does not reveal the values after completing the commit phase correctly, they will lose their deposit. However, the participant might benefit from a biased output of the random beacon in the downstream application, for which they might be willing to cheat even at the cost of losing the deposit. In order to prevent this possibility of adversarial manipulation on the game result, we make use of a verifiable delay function $\mathrm{VDF}(\cdot)$ as in~\citep{boneh:vdf} and require the participant to provide all necessary parameters for the evaluation of the VDF as part of the commit message. We then check that the provided VDF really evaluates to the strategy $s_i$ of the player and otherwise, simply remove player $i$ from the game. Of course, the VDF evaluation time should be long enough to ensure it cannot be evaluated until the reveal phase is over.

Using this technique, an adversary can have no information about the final output by the end of the reveal phase. Therefore, for all participants, revealing values honestly is always a strictly better strategy than not revealing values. The game is then executed when all the values are revealed and all the VDFs are evaluated and it is ensured that cheating participants are excluded with their deposits confiscated.

\paragraph{Avoiding the Random Oracle Assumption} Note that, even if $v$ is sampled from a uniform distribution, the output of a VDF on $v$ is not guaranteed to be uniformly random. In existing constructions of random beacons that rely on VDFs, a hash function is usually applied to the output of VDF to get random numbers. This is based on the classical random oracle assumption. However, using our RIG game, we have a more straightforward way of providing a game-theoretic guarantee of uniform randomness without relying on the random oracle assumption.

\paragraph{Bitwise Cutting} We can guarantee the delivery of uniformly-random output values by defining the final output  as $\tilde{v}=v_1 + \mathrm{VDF}(v_2)$, where $v_1$ contains the bits on the right half of $v$ and $v_2$ the bits in its left half. Since $v$ is uniformly random, then $v_1$ and $v_2$ are independent uniformly-random integers. Whatever distribution $\mathrm{VDF}(v_2)$ has, the sum $\tilde{v}$ would still be uniformly distributed.

Finally, note that this approach works as long as at least one of the participants is honest. So, in a proof-of-stake scenario, if we choose $n$ participants for each random number generation, we need to ensure that honest miners have much more than $1/n$ fraction of the stake in the cryptocurrency, so as to ensure that at least one honest participant is chosen with high probability. We also assume that at least one participant is reliable, but this is already incentivized by our game.

\begin{figure}[H]
    \centering
    \includegraphics[width=0.95\textwidth]{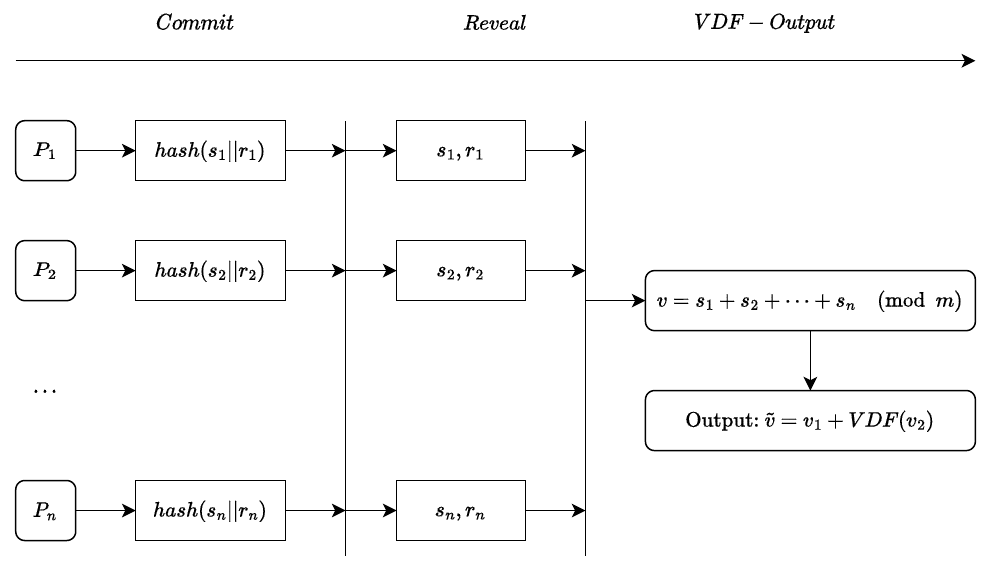}
    \caption{RIG Implemented using a Commitment Scheme and VDF}
    \label{fig:rig-commit}
\end{figure}

\paragraph{Diagram} Figure \ref{fig:rig-commit} shows the procedures of RIG implemented using a commitment scheme and VDF. In the commit phase, every participant sends the hash commitment. In the reveal phase, everyone is supposed to reveal her choice of the value $s_i$ and random nonce $r_i$. Finally $VDF$ is applied to the left half bits $v_2$ of the modular sum $v$, to yield the output random number $\tilde{v}=v_1+\text{VDF}(v_2)$.

\section{Random Beacon using Publicly-Verifiable Secret Sharing}\label{subsection:rig-pvss}

The fundamental drawback of relying on commitment schemes in random beacons is the possibility of adversarial manipulation in the reveal phase by not revealing values. We have already seen a solution using VDFs. A publicly-verifiable secret sharing scheme solves the same issue differently, i.e.~by forcing the opening of commitments, at the cost of increased communication complexity. In this chapter, we follow the PVSS constructions in~\citep{berry:simple-pvss}. 

An execution of the RIG game in a PVSS scheme consists of three phases: prepare, distribute and reconstruct. 

\paragraph{Prepare Phase} In the prepare phase, all eligible participants inform each other that they are selected. Under a synchronous communication setting, all honest participants can reach a consensus on the list of participants. More specifically, we assume a blockchain protocol that a transaction will be added to a finalized block within known bounded time slots after it is broadcasted. Each participant firstly broadcasts a signed prepare message $(ss\_id, proof_i)_i$ to announce their identity along with eligibility proof for the current session of execution. By the end of prepare phase, all prepare messages should be included in the blockchain and are synchronized across all nodes. Suppose the list of participants is $\{P_i\}_{i=1}^n$. 

In the distribute and reconstruct phases, each participant $P_i$ runs a PVSS scheme to share their value $s_i$ to the other $n-1$ participants. This is exactly as described in Section~\ref{sec:pvss}. 

\paragraph{Distribute Phase} In the distribute phase, every participant should send valid $(n-1, t)$-threshold secret shares to others along with a proof of commitment and consistency. The shares are publicly verifiable so that a dishonest participant who distributes invalid shares can be discovered and excluded from the game. Hence, by the end of distribute phase, all honest participants release their correct shares and receive correct shares from other honest participants. If a dishonest participant distributes some invalid shares or does not distribute part of the shares, they will be discovered and deleted from the list of participants. As long as the number of dishonest participants is less than $t$, they cannot decrypt any secret from honest participants in the distribute phase. 

\paragraph{Reconstruct Phase} In the reconstruct phase, each participant can reveal their value and share the decrypted secret shares they received. If the number of honest participants is at least $t$, then the pooling mechanism is successful and anyone can find all the secrets from valid secret shares, without the help of any dishonest participant. The dishonest participants cannot mislead honest participants by providing wrong decryption of secret shares in reconstruction, because the decryption for reconstruction also requires a publicly verifiable proof.  

\begin{figure}[H]
    \centering
    \includegraphics[width=0.95\textwidth]{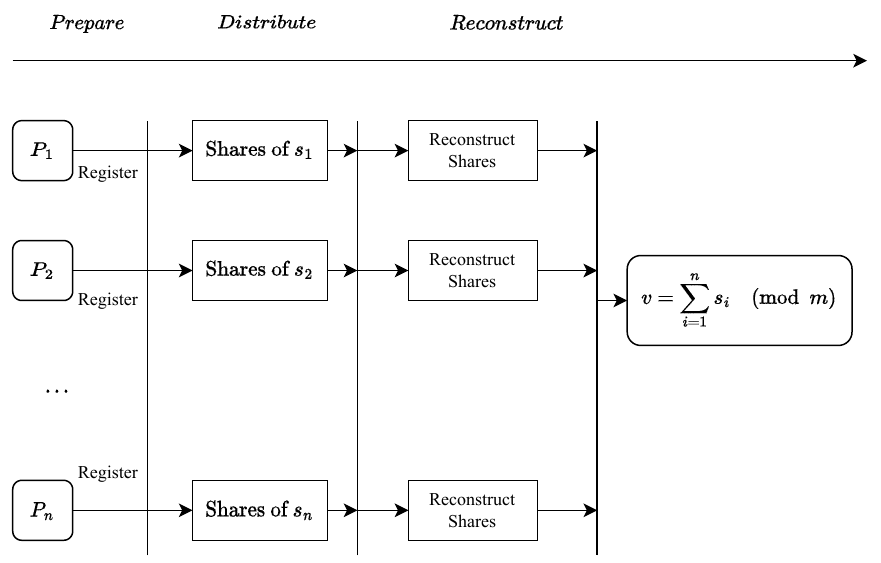}
    \caption{RIG Implemented using a PVSS Scheme}
    \label{fig:rig-pvss}
\end{figure}

\paragraph{Diagram} Figure \ref{fig:rig-pvss} illustrates RIG implemnted using a PVSS scheme. In the prepare phase, eligible parties register and they will know the list of all participants. Then each participant $P_i$ chooses a value $s_i$ and publishes its shares among the network. After enough participants send reconstruct shares, the output random number can be computed.

\paragraph{Analysis and Summary of PVSS in PoS} Suppose the number of dishonest participants is $f$. PVSS requires $f < t \le n-f$. Therefore, we can assume that $n\ge 2 \cdot f+1$ and let $t = \lceil n/2 \rceil$. In other words, we need to assume that more than half of the participants are honest. In a proof-of-stake application, the set of participants for each execution are sampled from the population of miners based on their stake. If we want $n\ge 2 \cdot f+1$ to hold with overwhelming probability, then the dishonest stake ratio should be strictly smaller than $1/2$ and preferably much smaller. Moreover, $n$ should not be too small. In other words, this approach assumes that most of the stake, strictly more than half and preferably much more than half, is owned by honest participants. This is in contrast to the previous approach that only needed more than $1/n.$

\section{Further Details of the Approach}\label{subsection:rig-other-parts}
\paragraph{Participant Selection Rules} Proof-of-stake blockchain protocols are important applications of random beacons. To prevent Sybil attacks and enforce proof-of-stake, it is common to sample a small subset of participants based on stake ratios for the random beacon. Verifiable random functions (VRFs)~\citep{micali:vrf} are popular for the purpose of selecting participants. A VRF requires a random seed, which can be the output of an RIG game in the previous round. Similar to the treatment for VDF outputs to ensure a uniform distribution, we can also use the trick of separating the bits of the random seed $v$ to two parts $v_{1}$ and $v_{2}$ and using $v_1 + \text{VRF}(v_2)$ instead of $\text{VRF}(v)$. 

\paragraph{Proposed Sorting Rule} In contrast to RANDAO and many other random number generators, our RIG game is sensitive to the order of participants. The result of the RIG game is not only the output value, which is the sum of all valid values submitted by the participants, but also the payoffs. The honest participants who reveal their values faithfully might receive different rewards/penalties depending on the ordering of participants. However, this is not a limitation. As before, we can use the output of the previous RIG round to generate a random ordering for the current round. 

\paragraph{Ensuring an Even Number of Participants} The RIG game requires an even number of participants, so if the number of valid participants is odd, we will remove one participant arbitrarily. To make sure this does not have an effect on consensus, we can remove the participant for whom $h(\texttt{commit message})$ has the largest value.

\paragraph{Alternative Sorting Rules} If relying on a previous round of RIG is undesirable, or if a previous round does not exist, we can also consider using the following sorting rules:  
\begin{itemize}
\item Order by $hash(ss\_id, pk_i)$ for participant $P_i$ with public key $pk_i$. Using $hash(pk_i)$ alone might give some participants permanent advantages/disadvantages. 
\item The order of commit/distribute phase transactions on blockchain. Note that we should not use the order of reveal/reconstruct phase transactions, because the game result in terms of reward/penalty can be manipulated. 
\end{itemize}

\paragraph{Deposits and Outputs} Every participant puts down a deposit $d$ at the same time they send their commit message. The value of $d$ is fixed by the protocol based on the needs of the particular downstream application. After collecting all the valid values $s_i$ and ordering of the participants $P_i$, $i\in [1, n],$ the result has the format $(v, \{P_i, u_i\})$, where $u_i$ is the payoff of participant $P_i$. The values $s, s_i$ are in $\{0, 1, \dots, m\}$, where $m$ is a parameter of RIG game, i.e.~the number of strategies for each player. The output random number is computed as $v = \sum_{i=1}^n s_i \pmod m$. If VDF is applied to $v$, the output random number is $\tilde{v}=v_1+\text{VDF}(v_2)$ as defined previously. Note that all dishonest players are excluded from the sum. If a player does not release their value or otherwise cheats, then they will be punished by confiscating their deposit and they will not be included in the game.

\paragraph{Design of Incentives} Each  honest participant $P_i$ receives a payoff of the form $rw_i = u_i(s_1, \ldots, s_n) + c$. Recall that $u_i(s_1, \ldots, s_n)$ is the payoff of $P_i$ defined by the game matrix, which sums up to 0 among valid participants. The number $c$ is a constant defined by the specific application.  Generally, $c$ should be  positive and high enough to motivate honest participants to join the RIG game and perform its steps. When we require the participants to use blockchain transactions for communication, $c$ should at least cover the transaction fees. 
The deposit amount $d$ should also be larger than any reward that a participant can possibly obtain in the game in order to discourage dishonest behavior. 

\section{Assumptions and Limits to Applicability} \label{subsection:rig-assumptions}

In this section, we review the assumptions we made in our approach. If any of these assumptions are not met, then our RIG game protocol above is not applicable.

\paragraph{Network Assumptions} The most important assumptions are the network assumptions. Our RIG game relies on a synchronous communication network. All real-world blockchain networks use the internet and are effectively synchronous.

\paragraph{$\delta$-synchrony}
A broadcast network is $\delta$-synchronous if a message broadcasted by some user at time $t$ will be received by all other users by time $t+\delta$. 

When applied to blockchains, blockchain consensus protocols guarantee public ledgers with different levels of synchrony. In this paper, we rely on blockchain consensus protocols to achieve a synchronized view of RIG execution. In detail, we require $\Delta$-synchrony for blockchains: 

\paragraph{$\Delta$-synchronous blockchains}
A blockchain is $\Delta$-synchronous if any valid transaction broadcasted by some user at time $t$ will become part of the stable chain at time $t+\Delta$ in the view of all honest nodes. 

The $\Delta$-synchrony can also be described by \textit{persistence} and \textit{liveness} properties, which are widely guaranteed with different parameters in the blockchain community:
\begin{itemize}
\item Persistence with parameter $k$ means a transaction is declared stable if and only if it is in a block that is more than $k$ blocks deep in the ledger. 
\item Liveness with parameter $u$ means if all honest nodes in the system attempt to include a certain transaction, then after $u$ slots, all nodes report the transaction as stable.  
\end{itemize}

We assume that $\Delta$ is known to all nodes and use it to design the duration of commitment scheme approach and PVSS approach. Specifically, in the commitment scheme approach we must have: (i)~$T_{commit}, T_{reveal} > \Delta$; and (ii)~$T_{wait} > T_{Eval}$. After the reveal phase ends, each participant requires extra $T_{wait}$ time slots to compute the final output. 
Similarly, in the PVSS approach we must have 
$T_{prepare}, T_{distribute}, T_{reconstruct} >\Delta$. 

If the PVSS approach is implemented using off-chain communication, then we can design durations in terms of $\delta$. In any case, the approach will not work if the network/blockchain is not synchronous or if the time limits are too tight and messages are not guaranteed to be delivered before the beginning of the next phase.

\paragraph{Rationality Assumption} We proved that any rational party or parties, that are interested only in maximizing their own payoff, would play uniformly at random in an RIG game and would therefore be reliable. This is because playing uniformly at random is the only alliance-resistant equilibrium in the game. Moreover, the uniformity of the output random number depends on having at least one reliable player. Therefore, we must assume that at least one player is rational and our approach would not work if none of the players are rational. However, this case is unlikely to happen in practice as we would normally expect all parties to be rational. 

\paragraph{Computation and Communication Complexity}
Our RIG can be implemented using a commit-reveal scheme or a PVSS scheme. Depending on the assumptions such as availability of reliable communication channels, the complexity might be different. Overall, we believe that the computation and communication complexity of RIG is better or comparable with existing efficient random number generation protocols and it imposes negligible overhead in applications under reasonable assumptions, as exemplified in Chapter~\ref{section:rig-in-blockchains}. In the presence of an underlying blockchain protocol that offers a decentralized ledger with reasonable synchrony ($\Delta$-synchrony), the commitment schemes require each participant of RIG to broadcast $2$ transactions of constant size on the blockchain. For PVSS schemes, each participant should also broadcast $2$ transactions on the blockchain but each transaction is of size $\Theta(n)$, sending shares to every other participant.   

\paragraph{Assumptions in Commitment Schemes} The commitment scheme approach inherits the standard cryptographic assumptions on cryptographic hash functions and verifiable delay functions.  

\paragraph{Assumptions in PVSS} The PVSS approach inherits the standard assumptions of PVSS schemes. Specifically, it requires that the probability of sampling a set of participants where $n < 2 \cdot f + 1,$ i.e.~a set in which honest participants are not the strict majority, is negligible. 

%% file: chapter5.tex
\chapter{RIG in Proof-of-Stake Protocols} \label{section:rig-in-blockchains}

We now show how our RIG random beacon can supplant standard PoS protocols. In general, the RIG random beacon, be it implemented by the commitment scheme approach or the PVSS approach, is applicable to any PoS protocol that requires an evolving random seed to select miners. Overall, using our RIG as the source of randomness only introduces negligible overhead in terms of transaction throughput. 

\paragraph{Slots and Epochs} If we use the RIG random beacon to generate the random seed in a proof-of-stake protocol, a single execution of the RIG game updates the random seed once. Usually, a single execution spans an epoch, which consists of multiple slots where the same random seed is repeatedly used. The blockchain protocol is modified to consecutively run the RIG random beacon to update the random seed in every epoch. The participants of RIG random beacon of each epoch are randomly selected, e.g.~based on the RIG result of the previous epoch. Note that our approach can also be applied for every block, instead of every epoch, but this would require more communication. 

\section{RIG in Ouroboros Praos}\label{subsection:rig-in-ouroboros-praos}
\paragraph{Ouroboros Praos} Ouroboros Praos is the second generation of proof-of-stake protocols in the Ouroboros family and the underlying protocol of Cardano cryptocurrency~\citep{bernardo:ouroboros-praos}. The selection rule for random seed generation participants in Ouroboros Praos is based on a VRF. The random beacon concatenates the VRF output of the participants and applies a random oracle hash function on the concatenated output. Each participant is also a miner and announces their VRF output along with their new block. The generated random seed is used in the next epoch, which consists of a number of slots. The protocol waits for enough slots until the seed generation is synchronized among all participants for the next epoch. 

\begin{figure}[H]
    \centering
    \includegraphics[width=0.95\textwidth]{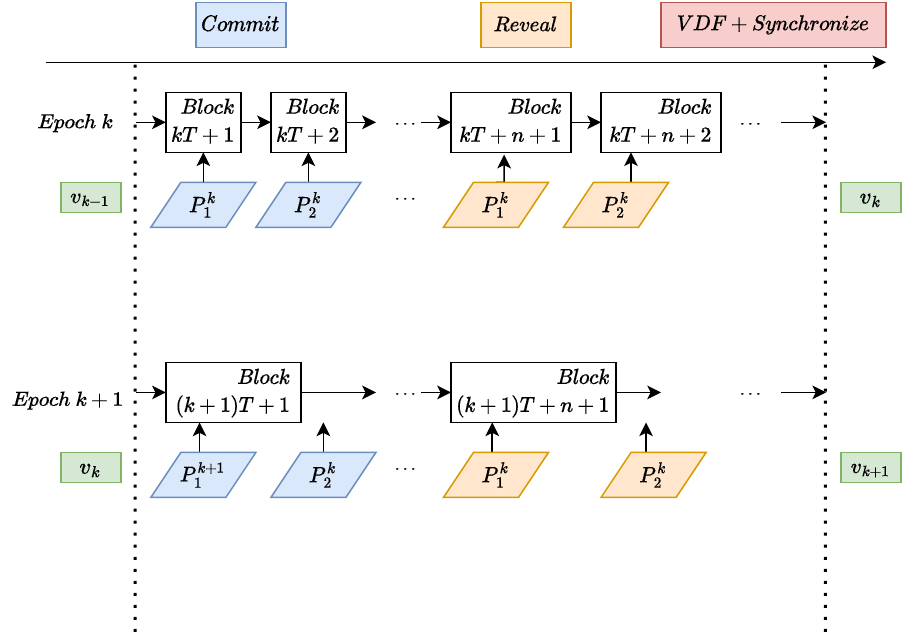}
    \caption{RIG in Ouroboros (using a Commitment Scheme)}
    \label{fig:rig-ouroboros}
\end{figure}

\paragraph{RIG Configurations in Ouroboros Praos} We can substitute the random beacon of Ouroboros Praos with our RIG. This is feasible because Ouroboros Praos assumes $> 50$\% honesty and suitable synchronous networks. If we use the commitment scheme approach, we can reuse the VRF selection rule and epoch/slot timing system. Figure \ref{fig:rig-ouroboros} shows how RIG is executed in Ouroboros Praos. The major difference is that the execution of RIG consists of two phases of communication. Therefore, it requires $T_{reveal} + T_{wait}$ more slots within an epoch to reach a consensus on the result of RIG. Besides, we improve the VRF selection using the bitwise split trick to ensure a uniform distribution, instead of pseudo-random sampling. In Cardano, an epoch lasts for $5$ days, and the transaction confirmation time is around twenty minutes. When using the commitment scheme, we require more time ($\geq$ 3$\times$ transaction confirmation time) within an epoch for the extra reveal phase and VDF computation time to reach a consensus on the result of RIG, which is negligible. 

\section{RIG in Algorand}\label{subsection:rig-in-algorand}

\paragraph{Algorand} It is also possible to integrate our RIG in Algorand~\citep{micali:algorand}. Algorand executes a Byzantine agreement protocol to achieve no-fork and one-minute-liveness properties, which guarantee a one-minute-synchronous blockchain for the RIG random beacon. Algorand assumes that honest users hold more than two-thirds of the total stake, which satisfies the requirement of the PVSS implementation of RIG. 

\begin{figure}[H]
    \centering
    \includegraphics[width=0.95\textwidth]{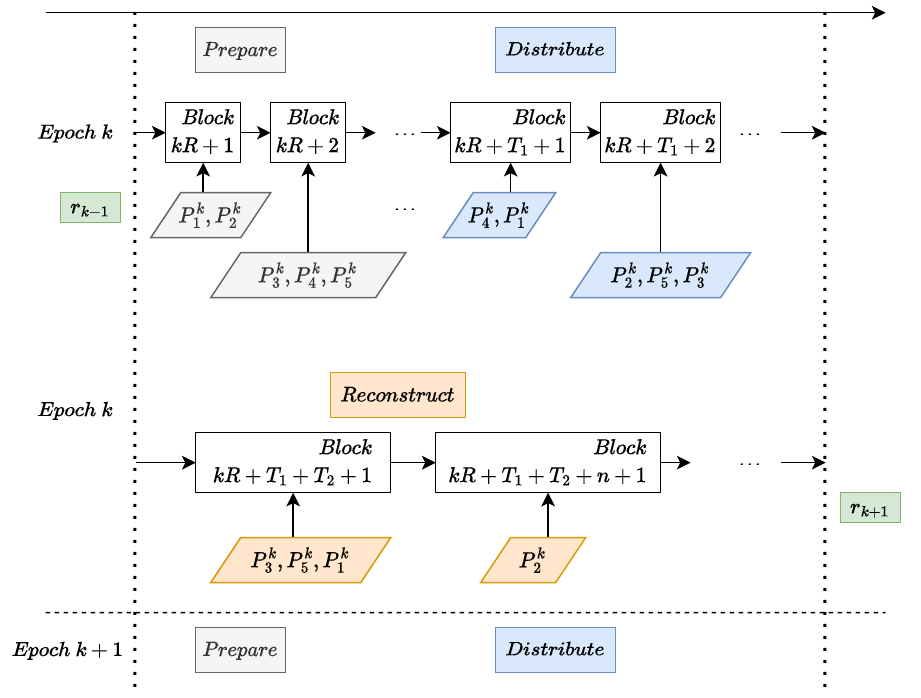}
    \caption{RIG in Algorand (using a PVSS Scheme)}
    \label{fig:rig-algorand}
\end{figure}

\paragraph{RIG Configurations in Algorand} Algorand requires an evolving random seed for VRF-based selection of miners and Byzantine agreement protocols. The random seed is updated every $R$ rounds by applying the VRF of the current miner to the previous random seed and the current epoch number. This is again pseudo-random and the output might deviate from uniform distribution. We can use our RIG random beacon to generate the random seeds for Algorand. If we use on-chain communication in the PVSS approach, then we have to select the participants of RIG separately from the committee for each single round, because we want the participants of RIG to be active for multiple slots ($T_{prepare} + T_{distribute} + T_{reconstruct}$ slots) without obstructing the growth of blockchain. Each single execution of the RIG spans the duration of an epoch, which consists of $R$ rounds. Algorand reaches consensus within $1$ round, and updates the random seed once every 1000 rounds, which is sufficient for a PVSS execution. Moreover, assuming that $R=1000$ and 100 participants join the RIG game, using RIG decreases the transaction throughput by less than $1\%$. Figure \ref{fig:rig-algorand} illustrates how RIG is integrated into Algorand. In the Distribute and Reconstruct stages, participants can send the desirable messages in arbitrary order, as long as they do it during the right time slots. $v_{k-1}$ is a random seed from the previous epoch and used in the current epoch $k$. Within $<R$ slots, the RIG participants can generate and reach consensus on a fresh random seed $v_{k}$ for future epochs. In this example, $P_4^k$ does not send a \textit{Reconstruct} transaction, but her value can be reconstructed by other honest participants. 

\paragraph{Comparison} The execution of the RIG random beacon is parallel to other parts of Algorand. While other parts require an evolving random seed from the RIG random beacon, they can reuse the previous random seed until a new seed is computed and synchronized. Compared to the random seed updating procedure in Algorand, our RIG random beacon is bias-resistant and rules out the possibility of adversarial manipulation on the random seed assuming the selected participants satisfy the assumption of honest majority. 

%% file: chapter6.tex
\chapter{Conclusion}\label{section:conclusion}

In this thesis, we presented a game-theoretic beacon for distributed random number generation. We showed that our approach is bias-resistant, unpredictable, available, verifiable and incentivizes every participant to be reliable, i.e.~faithfully provide a uniform random input. Even if only one of the participants is rational and thus reliable, the output number is guaranteed to be unbiased. Additionally, our approach does not use pseudo-randomness at any point and instead only relies on well-incentivized game-theoretic randomness. Finally, even though the approach is general and not limited to blockchain use-cases, we showed that one can easily augment common proof-of-stake protocols to include our randomness beacon for the task of selecting their miners. This ensures that proof-of-stake protocols choose the miners fairly, i.e.~exactly in proportion to their stake.

For the future work, we aim at a random integer protocol that keeps each participant's value secret throughout the execution of the protocol, probably by relying on homomorphic encryption~\cite{DBLP:conf/stoc/GoldwasserM82,DBLP:phd/us/Gentry09}, zero-knowledge proofs~\cite{DBLP:conf/crypto/BitanskyC12,DBLP:conf/icalp/Ben-SassonBHR18,DBLP:conf/sp/BunzBBPWM18,DBLP:conf/crypto/Ben-SassonBHR19} or alternative value aggregation methods such as~\cite{DBLP:journals/ijcomsys/LiuLCCJ14}. 

%% file: reference.tex
\bibliography{reference}

%% file: publication.tex
\null\skip0.2in
\begin{center}
{\bf \Large \underline{List of Publications}}
\end{center}
\vspace{12mm}

\noindent [1] \textbf{Z. Cai}, and A. Goharshady, ``Trustless and Bias-resistant Game-theoretic Distributed Randomness,'' in \textit{IEEE ICBC}, 2023. 

\noindent [2] \textbf{Z. Cai}, and A. Goharshady, ``Game-theoretic Randomness for Proof-of-Stake,'' in \textit{MARBLE}, 2023. 

\noindent [3] \textbf{Z. Cai}, S. Farokhnia, A. Goharshady, and S. Hitarth, ``Asparagus: Automated Synthesis of Parametric Gas             Upper-bounds for Smart Contracts,'' in \textit{OOPSLA}, 2023. 